\documentclass[aps,pra,twocolumn,nofootinbib
,superscriptaddress
,10pt,showpacs]{revtex4-1}
\def\Title {Bose-Einstein condensation in a minimal inhomogeneous system}

\newcommand{\xpct}[1]{\bigl\langle #1 \bigr\rangle}
\newcommand{\V}[1]{\mathbf {#1}}

\newenvironment{smallpmatrix}{\left(\begin{smallmatrix}}%
{\end{smallmatrix}\right)}

\usepackage[utf8]{inputenc}
\usepackage{prettyref}
	\newcommand{\pr}[1]{\prettyref{#1}}
	\newrefformat{app}{appendix~\ref{#1}}
	\newrefformat{eqn}{Eqn.~(\ref{#1})}
	\newrefformat{eqns}{Eqns.~(\ref{#1})}
	\newrefformat{sec}{section~\ref{#1}}
	\newrefformat{subsec}{section~\ref{#1}}
	\newrefformat{subsubsec}{section~\ref{#1}}
	\newrefformat{fig}{Fig.~\ref{#1}}

\usepackage[pdftex]{graphicx}
\usepackage{grffile}	
\usepackage{subfigure}
\usepackage{amsmath,amsfonts,amssymb}
\usepackage{dsfont}  

\usepackage{enumitem}
\setitemize{topsep=2pt, itemsep=0pt, parsep=2pt, leftmargin=1.5em}
\setenumerate{topsep=2pt, itemsep=0pt, parsep=2pt, leftmargin=1.5em}

\allowdisplaybreaks[1] 

\usepackage[pdftex, bookmarks, bookmarksopen,
bookmarksopenlevel=1, bookmarksnumbered=true,
pdftitle={\Title},colorlinks, linkcolor=blue, urlcolor=blue]{hyperref}	

\usepackage[american]{babel}

\begin{document}
\title{\Title}
\author{Christopher Gaul}
\affiliation{Departamento de F\'isica de Materiales, Universidad Complutense, E-28040 Madrid, Spain}
\affiliation{GISC and CEI Campus Moncloa, UCM-UPM, E-28040 Madrid, Spain}
\author{Jürgen Schiefele}
\affiliation{Departamento de F\'isica de Materiales, Universidad Complutense, E-28040 Madrid, Spain}
\date{May 17, 2013} 
\begin{abstract}
We study the effects of repulsive interaction and disorder on Bosons in a two-site Bose-Hubbard system, 
which provides a simple model of the \emph{dirty boson problem}. 
By comparison with exact numerical results, we demonstrate how a straightforward application
of the Bogoliubov approximation fails even to deliver a qualitatively correct picture:
It wrongly predicts an increase of the condensate depletion due to disorder.
We show that, in the presence of disorder,  the   noncommutative character of the condensate operator has to be retained
for a correct description of the system.
\end{abstract}
\maketitle
%
%
%
%
%
%
%
%
%
%
\section{Introduction}
The interplay of disorder and interaction in bosonic systems, known as the
\emph{dirty boson problem} \cite{Fisher1989}, is responsible for the
superfluid--insulator transition observed in many condensed-matter systems,
like superfluid helium adsorbed on porous media~\cite{Crooker1983,*Wong1990}, high-$T_c$
superconductors \cite{Dubi_2007}, and light propagating in disordered
media~\cite{Schwartz_2007}.  While disorder, giving rise to Anderson
localization \cite{Anderson1958}, can destroy the superfluid condensate  and
take the system to a Bose-glass phase \cite{Fallani2007,Yu_2012}, weak repulsive
interactions have instead a delocalizing effect.  This competition has recently
been studied experimentally with Bose gases of cold atoms in optical lattice
potentials, where both the strength of  interaction and disorder can be
controlled experimentally \cite{White2009,Deissler_2010}.

\begin{figure}[bt]
 \includegraphics[width=0.85\linewidth,clip]{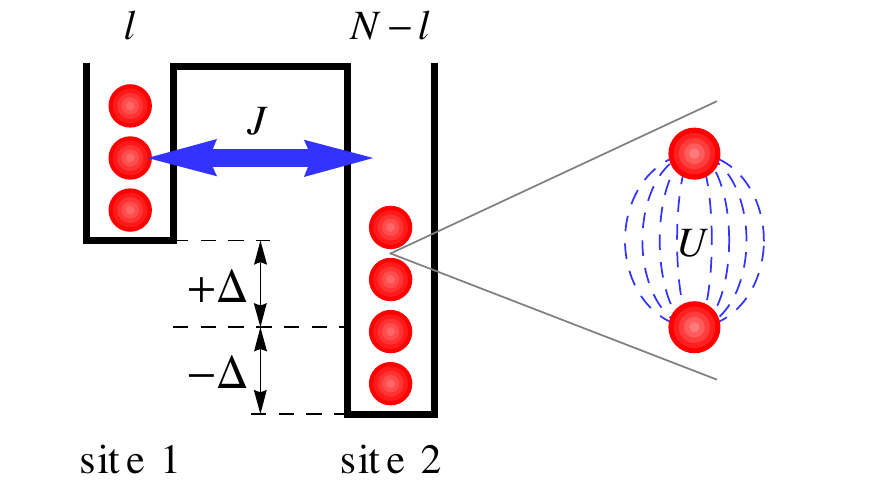}
 \caption{
 \label{fig:sketch}
 (Color online) The two-site Bose-Hubbard model described by Eqn.~(\ref{eqHamiltonian}).
 $J$, $U$, and  $\Delta$ denote the hopping amplitude, on-site interaction, and tilt, respectively.
 }
\end{figure}

The aim of the present work is to study the dirty boson problem.
Concretely, we seek to obtain more insight in the intricacies of the Bogoliubov
approximation  in the inhomogeneous case \cite{Bogoliubov1947,Gaul2011_bogoliubov_long,Muller2012_momdis}.
To this end, we consider a minimal model of  interacting Bosons in a system  
of only two lattice sites, 
described by a  Bose-Hubbard model
(also known as Josephson junction or Lipkin-Meshkov-Glick
model \cite{Lipkin1965,*Meshkov1965,*Glick1965}) with the Hamiltonian 
\begin{align}\label{eqHamiltonian}
 \hat H = -J (\hat a_1^\dagger \hat a_2 +\hat a_2^\dagger \hat a_1) + \Delta (\hat n_1 - \hat n_2) 
+ \frac{U}{2} (\hat n_1^2 + \hat n_2^2) ,
\end{align}
(see Figure~\ref{fig:sketch}).
Here, $\hat a_j$ denotes bosonic operators, $\hat n_j = \hat a_j^\dagger \hat a_j$,
and the parameters $J$, $U$, and  $\Delta$ are hopping amplitude, on-site interaction, and tilt (or energy mismatch between the sites), respectively.
In our toy model, the tilt  $\Delta$  represents disorder \cite{Zhou2010}.

The Bogoliubov approximation is an efficient method for the perturbative treatment of weakly interacting Bose condensates; 
it brings the Hamiltonian to a form quadratic in quasi-particle operators \cite{Bogoliubov1947}.  
%
These describe quantum fluctuations on top of the macroscopically occupied condensate
mode. 
The Bogoliubov excitations can be associated with the Goldstone mode of the system due to
spontaneously broken $U(1)$ symmetry \cite{Leggett1991}.  
The Bogoliubov approximation is valid for systems that approach a thermodynamic limit
such that both the particle number and the volume of the system tend  to infinity,
while the ratio of the two remains a finite constant. 
If the volume of the system is constrained \cite{Schiefele_2009},
additional finite-size effects play a role \cite{Fetter1972}.
We show below that for our model with only two lattice sites, 
a naive application of the Bogoliubov approximation even fails to deliver a qualitative description of the system in the presence of disorder.
Instead, we need to re-introduce the quantum character of the condensate mode 
to construct the $N$-particle wavefunction of the interacting groundstate.
In this way, we obtain results that agree with the exact diagonalization of Hamiltonian Eqn.~(\ref{eqHamiltonian}) in the
limit of large particle number.
In contrast to extended disordered systems \cite{Gaul2013,Muller2012_momdis}, we find that the tilt $\Delta$ counteracts the depletion of the condensate due
to interaction.


%
%
%
%
%
%
%
%
%
%
%
%
%
\section{Exact diagonalization}\label{sec:Exact}
We work with a fixed particle number $N$.
Then, the Hilbert space of \eqref{eqHamiltonian} is $N+1$-dimensional: $l$ particles on site 1 and $N-l$ particles on site 2, where $l$ runs from $0$ to $N$.
Numerically, it is straight forward to diagonalize the matrix
$H_{l l'} = \left\langle l \right| \hat H \left| l' \right\rangle$.
Throughout this work, we consider zero temperature,  so we take the eigenvector with the lowest energy and compute the one-body density matrix
\begin{equation}
\rho_{ij}=\xpct{\hat a_i^\dagger \hat a_j}
\end{equation}
According to Penrose and Onsager \cite{Penrose1956}, the condensate mode is identified as the eigenstate of $\rho$ with the largest eigenvalue $N_0$, and that eigenvalue is the population of the condensate.
Conversely, the depletion of the condensate $\delta N$ is the sum of all other eigenvalues. In the present case of only two sites, there is only the condensate and one other mode.
Examples of the $N$-particle ground-state wavefunction are shown in Figure \ref{fig_manybody},
results for the depletion in Figures \ref{fig:depletion_untilted} and \ref{fig:Condep-of-Delta}.
%
%
%
%
%
%
%
%
\section{Approximate analytical solution}
At temperatures well below the transition to the condensed phase,
it is convenient to separate the bosonic operators $\hat{a}_j$ 
into  condensate and  noncondensate part
\begin{align}
\hat{a}_j
	&=
	f_j \hat{a}_0
	+
	\delta \hat{a}_j
.
\label{eqn:a_def_full}
\end{align}
In our discrete two-site system, the numbers $f_j$ 
with the normalization
\begin{align}
|f_1|^2 + |f_2|^2 = 1
\label{eqTotalN}
\end{align}
are the analogue to the condensate wavefunction,
i.e.\ a macroscopic number of particles in a product state.
For the sake of simplicity, we will assume $f_{j}$ to be real in the following.

%
%
%
\subsection{Bogoliubov meanfield part}
\label{subsec:Bogo_mean}
We assume a large number of atoms on each of the two sites,
and continue by applying the Bogoliubov  approximation. 
It consists in substituting the operators $\hat{a}_0$ and $\hat{a}_0^\dagger$
with $\sqrt{N_0}$, where $N_0$ denotes the number of atoms in the condensed mode.
Further assuming a small condensate depletion with $N_0 \approx N$, we have 
\begin{equation}
\hat a_j 
	\approx \sqrt{N} f_j 
	+ 
	\delta \hat a_j 
.
\label{eqBogoliubov}
\end{equation}
With this approximation, we will first determine an approximate form
of the meanfield wave function $f_j$, 
and bring the 
Hamiltonian Eq.~(\ref{eqHamiltonian}) to a quadratic form in the fluctuation operators $\delta \hat a_j $.
However, in \pr{subsec:depletion},  we will show 
that for a correct description 
of the condensate depletion and the system's many-body wavefunction,
it is essential to re-introduce the noncommutative operator character of $\hat{a}_0$.

For technical reasons, we chose the grand canonical frame $\hat E = \hat H - \mu \hat N$, $\hat N = \hat n_1 + \hat n_2$.
We will always adjust the chemical potential $\mu$ as function of $J$, $U$, $\Delta$ and $N$ such that a given particle number $N$ is kept fixed.
The meanfield solution $f_j$ minimizes $E[\sqrt{N}f_j]$, 
i.e., the $f_j$ fulfill the the discrete Gross-Pitaevskii equation
\begin{subequations}\label{eqMFproblem}
\begin{align}
-J f_2 + (\Delta + U N f_1^2 - \mu) f_1 &= 0 \label{eqMFproblem1}\\
-J f_1 + (-\Delta + U N f_2^2 - \mu)f_2 &= 0. \label{eqMFproblem2}
\end{align}
\end{subequations}
Together with the constraint \eqref{eqTotalN}, the meanfield problem is fully
defined; $f_1$,  $f_2$, and $\mu$ are determined as functions of $J$, $\Delta$,
and $U N$.  
Introducing the population imbalance $n=N(f_1^2 - f_2^2)$ and 
writing $f_1^2 = \frac 1 2 (1+n/N)$ and $f_2^2 = \frac 1 2 (1-n/N)$,  $J$ can be eliminated 
from Eqs.\ \eqref{eqMFproblem1} and \eqref{eqMFproblem2}, and one finds
\begin{equation}\label{eqn:Imbalance_exact}
 \frac{n}{N} = \frac{\Delta}{\mu - U N} .
\end{equation}
With this, and by setting the determinant of the coefficient matrix of Eqs.\ \eqref{eqMFproblem1} and \eqref{eqMFproblem2} to zero, one finds the quartic equation for the chemical potential
%
\begin{align}\label{eqQuartic}
(X - UN/2)^2 (X^2 - J^2) - X^2 \Delta^2 = 0,
\end{align}
where $X = \mu - UN/2$.
To leading order in $\Delta$, this yields%
\begin{equation}
\label{muEq}
 \mu = \frac{UN}{2} -J  - \Delta^2 \frac{J}{2(J+UN/2)^2}   + o(\Delta^4).
\end{equation}
Note that the negative shift of the chemical potential due to the ``disorder'' $\Delta$ is analogous to Eq.\ (15) of \cite{Gaul2011_bogoliubov_long}.
Via \prettyref{eqn:Imbalance_exact}, the chemical potential determines the meanfield imbalance and the condensate wave function $f_j$.
%
%
%
%
%
%
%
%
\subsection{Bogoliubov noncondensate part}
\label{B_non_cond}
The meanfield wave function $f_j$ has been obtained from the minimization of
the meanfield energy functional. That means, the leading order of the relevant Hamiltonian $F =
\hat E [\hat a_j] - E[\sqrt{N}f_j]$ is quadratic in the quantum fluctuations:
\begin{align}\label{eqn:BgHamiltonian}
 \hat F 
  &= \frac{1}{2} \sum_{i,j}
  \bigl(\delta \hat a_i^\dagger,\delta \hat a_i \bigr)
  \begin{pmatrix}
   D_{ij}	& B_{ij} \\
   B_{ij}	& D_{ij}
  \end{pmatrix}
  \begin{pmatrix}
   \delta \hat a_j \\ \delta \hat a_j^\dagger
  \end{pmatrix} , \\
 D &= 
 \begin{pmatrix}
  2 U n_1 + \Delta -\mu	& -J \\
  -J			& 2 U n_2 - \Delta -\mu
 \end{pmatrix}, \quad
 B_{ij} = \delta_{ij} U n_j . \nonumber
\end{align}
Here, we find a typical feature of the Bogoliubov ansatz: 
\prettyref{eqn:BgHamiltonian} contains terms like $U n_1 \delta \hat a_1 \delta \hat a_1$, which destroy two particles, instead of destroying one particle and creating one particle. 
The particle number is not conserved and 
implicitly, we understand that missing particles have gone to the condensate mode.

In other words, the equations of motion mix creators and annihilators.
This can be resolved by the \emph{Bogoliubov transformation} to quasi-particles
\begin{align}
 \hat \beta_\nu = u^*_{\nu 1}\delta \hat a_1 + u^*_{\nu 2}\delta \hat a_2
                + v^*_{\nu 1}\delta \hat a_1^\dagger + v^*_{\nu 2}\delta \hat a^\dagger_2
.
\end{align}
%
Postulating $i \hbar \partial_t \hat \beta_\nu = [\hat\beta,\hat F] \stackrel{!}{=} \omega_\nu \hat \beta_\nu$ and a comparison of coefficients,
we arrive at the Bogoliubov-de-Gennes equations \cite{Pitaevskii2003} 
\begin{align}\label{eqBgEV}
 \sum_j
 \Bigl[
  \begin{pmatrix}
   D_{ij}	& -B_{ij} \\
   B_{ij}	& -D_{ij}
  \end{pmatrix}
 - \omega_\nu 
  \begin{pmatrix}
   \delta_{ij}	& 0 \\
   0	& \delta_{ij}
  \end{pmatrix}\Bigr] 
 \begin{pmatrix} u_{\nu j} \\
                 v_{\nu j}
 \end{pmatrix} = 0 .
\end{align}
As the matrix in \eqref{eqBgEV} is not Hermitian, we cannot expect the eigenvectors to be orthogonal. Rather, they fulfill the bi-orthogonality relation \cite{Fetter1972}
\begin{align}\label{eqBiOrt}
(\omega_\nu - \omega_\lambda^*) \sum_j(u_{\nu j}^* u_{\lambda j} - v_{\nu j}^* v_{\lambda j}) = 0 .
\end{align}
The matrix in \prettyref{eqn:BgHamiltonian} anticommutes with 
$\bigl(\begin{smallmatrix}
 0           & \delta_{ij} \\
 \delta_{ij} & 0
\end{smallmatrix}\bigr)
$.
So, if $(u_{\nu1},u_{\nu2},v_{\nu1},v_{\nu2})$ is an eigenvector with eigenvalue $\omega_\nu$, then $(v_{\nu1},v_{\nu2},u_{\nu1},u_{\nu2})$ is an eigenvector with eigenvalue $-\omega_\nu$, which simply corresponds to $\hat \beta_\nu^\dagger$. Thus, Bogoliubov modes occur in pairs.
%
%

A special mode $\nu=0$ is found by setting $u_{0j}=v_{0j}$. Then, Eq.\
\eqref{eqBgEV} becomes the discrete Gross-Pitaevskii equation
\eqref{eqMFproblem}, such that $u_{0j}=v_{0j} = f_j$ and $\omega_0 =
0$.
The corresponding operator $\hat \beta_0 =: \hat P$ is Hermitian. 
It can be interpreted as a kind of momentum associated to the Goldstone mode of the $U(1)$ symmetry breaking of Bose-Einstein condensation \cite{Lewenstein1996}.
There is a conjugate position $\hat Q$ satisfying $[\hat Q, \hat P]=i$.
Since only one regular mode remains, we drop the index $\nu=1$.
Both $\hat P$ and $\hat Q$ commute with $\hat \beta$ and $\hat \beta^\dagger$, and the operators
$\hat \beta$, $\hat \beta^\dagger$, $\hat P$, and $\hat Q$ form a complete set to express the $\delta \hat{a}_j$ and $\delta \hat{a}_j^\dagger$, such that the Bogoliubov Hamiltonian $\hat F$ reads
\begin{equation}
 \hat F = \omega ( \hat \beta^\dagger \hat \beta +{1}/{2} ) + \alpha \hat{P}^2/2 \label{eqn:BgHamiltonian_diagonal} .
\end{equation}
For $\Delta = 0$, one finds the usual Bogoliubov dispersion 
$\omega^{(0)} = \sqrt{2J(UN+2J)}$ and the inverse mass term $\alpha^{(0)} = UN$.
Both quantities are even functions of $\Delta$; the quadratic correction is calculated in the Appendix, Eqs.\ \eqref{eqn:2ndorder-w} and \eqref{eqn:2ndorder-alpha}.
%

The regular Bogoliubov mode is normalized as $\sum_j (|u_j|^2 - |v_j|^2) = 1$, such that $[\hat \beta,\hat\beta^\dagger]=1$.
%
%
%
%
%
%
%
\subsection{Many-body wavefunction and condensate depletion}
\label{subsec:depletion}
To construct an explicit expression for the many-body wavefunction, 
it is necessary to go back to the original definition (\ref{eqn:a_def_full}) of the field operator $\hat{a}_j$.
It then follows from the bosonic commutation relation $[\hat{a}_i,\hat{a}_j^\dagger]=\delta_{i j}$
that the operators $\delta \hat{a}_i$ and $\delta \hat{a}_i^\dagger$ obey the commutation relations 
\begin{align}
[\delta \hat{a}_i,\delta \hat{a}_j^\dagger]
	&=
	\delta_{i j} - f_i f_j^* [\hat{a}_0,\hat{a}_0^\dagger]
	=
	\delta_{i j} - f_i f_j^*
	\equiv
	\overline{\delta}_{i j}
,
\label{eqn:delta_def}
\end{align}
where the last equality defines the projection operator $\overline{\delta}_{i j}$.
Within the Bogoliubov approximation, that is, with  Eqn.~(\ref{eqBogoliubov}), the above relation would become 
$
[\delta \hat{a}_i,\delta \hat{a}_j^\dagger]
	\approx	
	\delta_{i j}
$.

In the ground state $|C_N\rangle$ of the noninteracting system,  all $N$ particles
occupy the condensate state, 
\begin{align}
|C_N\rangle
	&=
	\frac 1 {\sqrt{N!}}
	\bigl( \hat{a}_0^\dagger \bigr)^N
	|0\rangle
,
\nonumber
\end{align}
%
where $|0\rangle$ is the no-particle state or physical vacuum.
The effect of pairwise particle interaction is to deplete this condensate state, 
and thus the lowest state of the interacting system --the Bogoliubov vacuum denoted by $|\V0 \rangle$-- 
consists of a superposition of states, each  with a different number $p$ of  pairs of particles excited out of the condensate:
\begin{align}
|\V 0\rangle
	&=
	Z
	\sum_{p=0}^{N/2}\,
	(2^{p} p!)^{-1}
	\bigl( \delta\hat{a}_i^\dagger  A_{i j}  \delta\hat{a}_j^\dagger  \bigr)^p
	|C_{N-2p}\rangle
.
\label{eqn:Ansatz0}
\end{align}
(Summation over repeated indices is implied.)
The symmetric matrix $A_{i j}$ and the normalization constant $Z$ in the ansatz \pr{eqn:Ansatz0} can be determined
from the condition
$
\hat{\beta}|\V0\rangle
	= 0 
$.
We refer the reader to Ref.~\cite{Fetter1972} for details of the calculation.
With the abbreviations
$
\overline{u}_{i}^*
	\equiv
	\overline{\delta}_{i j}
	u_{j}^*
$,
$
\overline{v}_{i}^*
	\equiv
	\overline{\delta}_{i j}
	v_{j}^*
$, and
$
\overline{A}_{i j}
	\equiv
	\overline{\delta}_{i k}
	{A}_{k l}
	\overline{\delta}_{l j}
$,
it results in 
\begin{align}
\overline{A}_{i j}
	&=
	-
	\overline{v}_{i}^*
	\overline{v}_{j}^*/\beta
,
\label{eqn:A_def}
\end{align}
with 
$
\beta
	=
	\overline{u}_{1}^* \overline{v}_{1}^*
	+
	\overline{u}_{2}^* \overline{v}_{2}^*
$
and
\begin{align}
Z^{-2}
	&=
	\operatorname{exp}
	\biggl\{
	\sum_{p=1}^{N/2}
	\frac{\operatorname{Tr}[(\overline{A}^* \overline{A})^p]}{2 p}
	\biggr\}
.
\label{eqn:Z_def}
\end{align}
Eqn.~(\ref{eqn:Ansatz0}) together with Eqns.~(\ref{eqn:A_def}) and  (\ref{eqn:Z_def})
yield an explicit representation of the interacting ground state 
$|\V 0\rangle$.

\begin{figure}[bt]
 \includegraphics[width=0.85\linewidth,clip]{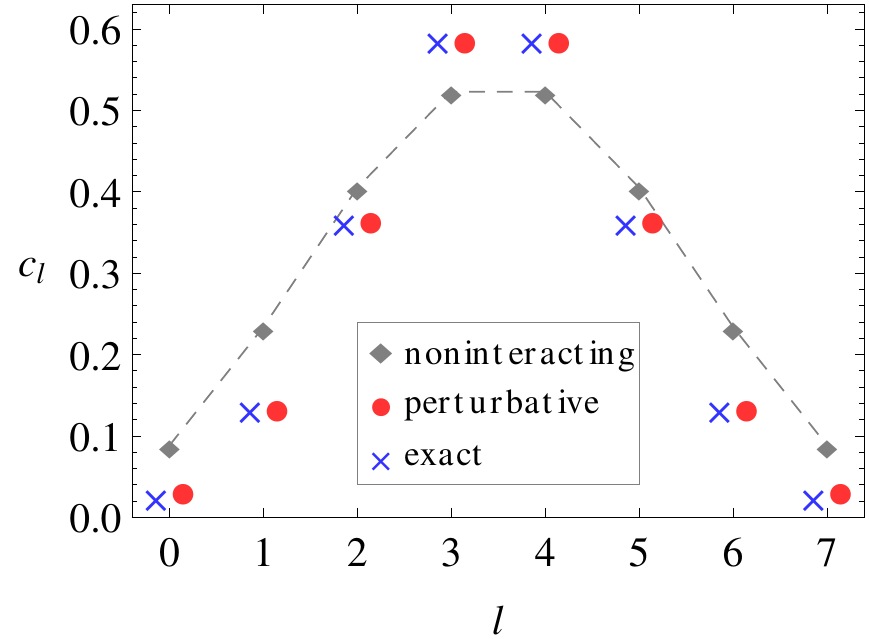}
 \caption{(Color online) Coefficients $c_l$ of the many-body wave function 
 $|\V 0\rangle=\sum_l c_l |l\rangle$ in the Fock basis for $\Delta=0$,
 $N=7$.  
 $|c_l|^2$ gives the probability to find $l$ particles on site~1.
 The gray diamonds connected by a dashed line show the noninteracting case of a
 pure condensate, where 
 $|c_l|^2 \propto \left(\begin{smallmatrix} N\\l \end{smallmatrix}\right)$, 
 which coincides with the exact-diagonalization results and the analytic results from
 Eqn.~(\ref{eqBgManyBodyCoefficients}) in absence of interaction.
 Blue  crosses and red dots show the exact numerical and the analytic results for
 $UN/J=2$, respectively (symbols slightly displaced
 horizontally).
 }\label{fig_manybody}
\end{figure}

Accordingly, the $N$-body wavefunction in configuration space can be written in the form
\begin{align}
\Psi (i_1, \dots, i_N)
	&\equiv
	N!^{-1/2} \langle 0 | \hat{a}_{i_1}, \dots, \hat{a}_{i_N}  | \V 0 \rangle
\nonumber
\\
	&=
	Z\,
	\sum_{p=0}^{N/2} \chi_p (i_1, \dots, i_N)
\;,
\label{eqn:psi_explicit}
\end{align}
where $\chi_p$ is the part of the wavefunction with exactly $p$ pairs of particles excited out of the condensate:
\begin{align}
&\chi_p  (i_1, \dots, i_N)
	=
	\biggl[ \frac{(N-2p)!}{N!}  \biggr]^{1/2}
\nonumber
\\
	&\times
	\sum \, \bigl[ 
	\overline{A}_{i_1 i_2}
	\dots
	\overline{A}_{i_{2 p-1} i_{2 p}}
	\times
	f_{i_{2p+1}}
	\dots
	f_{i_{N}}
	\bigr]
.
\label{eqn:chi_p_def}
\end{align}
For each pair of non-condensate particles occupying  the sites $i$ and $j$, 
there is a factor $\overline {A}_{i j}$ from \pr{eqn:Ansatz0}, for each condensate particle at site $i$ a factor $f_i$.
The sum in Eqn.~(\ref{eqn:chi_p_def}) runs over the $N![(N-2p)! \, p!  \, 2^p]^{-1}$ distinct ways of choosing $p$ different pairs 
from the $N$ variables $\{ i_1, \dots, i_N \}$.
With Eqns.~(\ref{eqn:psi_explicit})-(\ref{eqn:chi_p_def}), we obtain 
\begin{align}
\rho_{i j}
	&=
	N \sum_{i_2,\dots,i_N}
	\Psi(i_i, i_2, \dots , i_N)
	\Psi^*(i_j, i_2, \dots , i_N)
\nonumber
\\
	&=
	N_0 f_i f_j
	+
	\overline{v}_i\overline{v}_j^*
\label{eqn:rho}
\end{align}
for the one-body density matrix,
where $N_0 = N - |\overline{v}_1|^2 - |\overline{v}_2|^2$.

To compare the interacting ground state \prettyref{eqn:Ansatz0} with the results of exact diagonalization, 
we need to expand the wave function \eqref{eqn:psi_explicit} in the Fock basis ($l$ bosons on the left and $N-l$ bosons on the right site):
%
%
\begin{align}
 |\V0\rangle
  &=  \sum_{l=0}^N c_l |l\rangle , &
 c_l &= \begin{pmatrix}
        N\\
        l
       \end{pmatrix}^{\frac{1}{2}}
       \Psi(\underbrace{1 \ldots 1}_{l \text{ times}},2 \ldots 2) \, . 
\label{eqBgManyBodyCoefficients}
\end{align}
%
Figure \ref{fig_manybody} shows an example for $N=7$ particles, i.e., with $0 \leq p \leq3$ pairs in
\prettyref{eqn:psi_explicit}.
For moderate interaction $U \ll J$,
the agreement with data from the exact diagonalization 
is good despite of the small number of particles.  
Compared to the noninteracting case, the amplitudes for large $l$ and large $N-l$ are suppressed, i.e.,
the interacting system disfavors particles to cluster on one of the sites.
\begin{figure}[tb]
 \includegraphics[width=0.85\linewidth,clip]{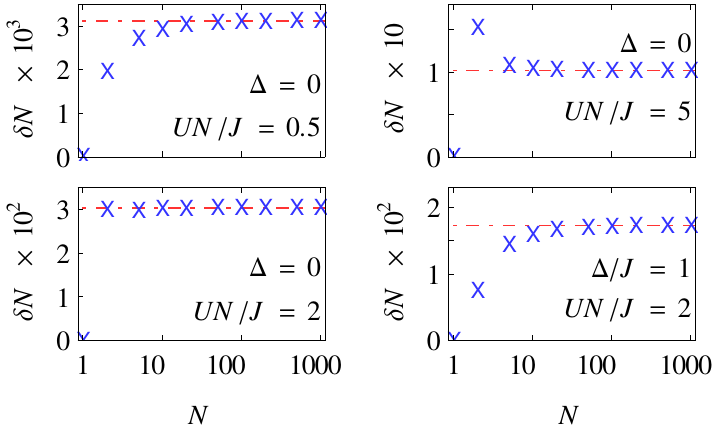}
\caption{%
(Color online) Number of non condensed particles $\delta N$ as a function of the total 
particle number $N$ for different values of $UN/J$ and $\Delta$.
Blue crosses show results from the exact diagonalization.
For large $N$ these points converge to the analytical result
given in Eqn.~\eqref{eqn:CondepFetter} (dashed red lines).
\label{fig:depletion_untilted}
}
\end{figure}

The density matrix Eqn.~(\ref{eqn:rho}) allows us to calculate the condensate depletion 
\begin{align}
\delta N 
	&= N-N_0 
	= 
	|\overline{v}_1|^2 + |\overline{v}_2|^2
.
\label{eqn:CondepFetter}
\end{align}
In Fig.~\ref{fig:depletion_untilted}, $\delta N $ is shown as a function of $N$.
For large $N$, the numeric results obtained by exact diagonalization converge to the value
given by Eqn.~(\ref{eqn:CondepFetter}).
Note that for $\Delta=0$,  we have $f_j = 1/\sqrt{2}$. Eqn.~(\ref{eqn:delta_def}) becomes
$\overline{\delta}=\bigl(\begin{smallmatrix}
\phantom{-}1 &          - 1 \\
         - 1 & \phantom{-}1
\end{smallmatrix}
\bigr)/2$, and with $v_1=-v_2$,
we arrive at $\overline{v}_j=v_j$.
Hence in this case,
\begin{equation}
\lim_{\Delta\to0}{\delta N} 
	= 
	|v_{1}|^2 + |v_{2}|^2
	\equiv
	\delta N_{\mathrm{Bg}} 
, 
\label{eqNaiveCondep}
\end{equation}
that is, the condensate depletion is correctly 
described within the simple Bogoliubov approximation of section~\ref{B_non_cond}.
\begin{figure}[tb]
 \includegraphics[width=0.85\linewidth,clip]{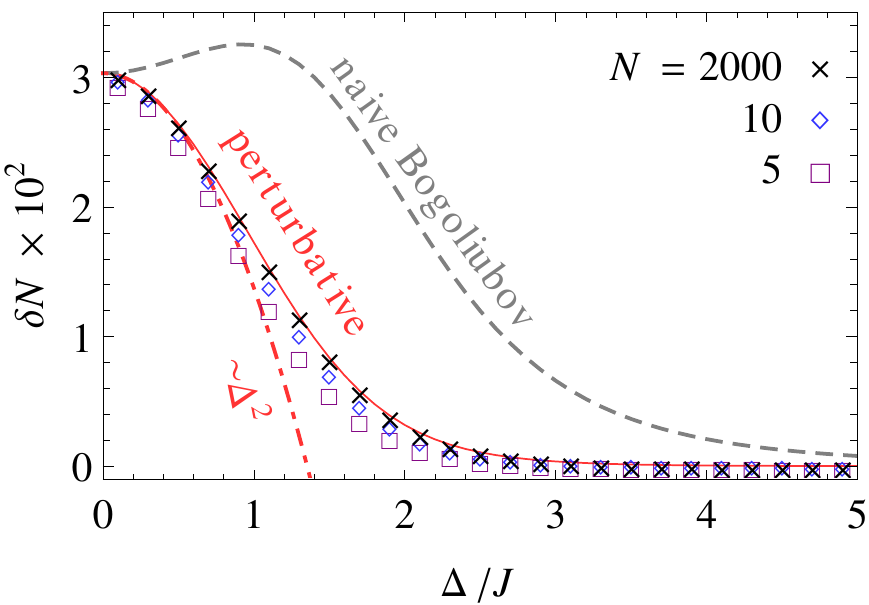}
\caption{(Color online) Condensate depletion as function of the tilt $\Delta$ for interaction
$UN=2J$. Symbols show results from the exact diagonalization for different particle numbers, 
lines show the analytical results of \pr{eqn:CondepFetter} and Eqn.~\eqref{eqNaiveCondep}.
\label{fig:Condep-of-Delta}}
\end{figure}
However, for $\Delta\neq0$, this is not the case:
Fig.~\ref{fig:Condep-of-Delta} shows  the condensate depletion as a function of the tilt $\Delta$. 
While $\delta N$ of Eqn.~(\ref{eqn:CondepFetter}) (red line) matches well with the 
results of the exact diagonalization (symbols), $\delta N_{\mathrm{Bg}}$ (gray dashed line) does not even qualitatively
match the functional form of $\delta N(\Delta)$,
and the Bogoliubov approximation $\hat{a}_0\to N_0$, resulting in 
$
[\delta \hat{a}_i,\delta \hat{a}_j^\dagger]
	=
	\overline{\delta}_{i j}
	\to \delta_{i j}
$
is not valid.
%
%
%
%
%
%
%
%
%
%
\section{Summary and discussion}
\label{sec:summary}
%
While either disorder or interaction alone tend to diminish the phase coherence 
in bosonic systems, the numerical calculations in Ref.~\cite{Zhou2010} found the --at first glance counterintuitive--
result that a combination of the two can actually enhance coherence.
In the present work, we observe this behaviour in the variation of the condensate depletion with the disorder
parameter $\Delta$. Figure~\ref{fig:Condep-of-Delta} shows how an increase in $\Delta$ restores  the condensate
population, counteracting its  depletion by repulsive interaction.

For the two-site system under consideration here, both condensate population and total particle number can be assumed large. 
However, a naive application of the Bogoliubov approximation is not valid, as a thermodynamic limit 
cannot be defined due to the fixed size of the system \cite{Fetter1972}.
By explicit construction of the  $N$-particle ground state, we
showed that the usual Bogoliubov wavefunctions $u$ and $v$ appear in a modified form $\overline{u}, \overline{v}$ in the one-body
density matrix of our system: they have to be corrected
by terms proportional to $[\hat{a}_0,\hat{a}_0^\dagger]$.  
In extended systems, these terms are multiplied with the inverse volume of the system,
which renders them negligible in the thermodynamic limit \cite{Fetter1972}.

Our comparison with exact numerical results reveals that the Bogoliubov description, which --by definition-- 
neglects the noncommutative character of 
the condensate operator $\hat{a}_0$,  
fails to describe the two-site system in the presence of disorder.
Therefore,  
a careful description of the interacting condensate particles is mandatory to 
capture the interplay between interaction and disorder within 
small inhomogeneous Bose systems.
%
%
%
%
%
%
\begin{acknowledgments}
We thank I. Zapata and H. Ranjan for useful comments.
Research of C.G. was funded by a PICATA postdoctoral fellowship from the Moncloa Campus of International Excellence (UCM-UPM) and supported by MINECO (MAT 2010-17180).
J.S. acknowledges financial support from the Marie Curie ITN \emph{NanoCTM}. 
\end{acknowledgments}
%
%
%
%
%
%
\appendix
%
%
%
%
%
%
\section{Analytical solution of the Bogoliubov-de-Gennes equation}
\newcommand{\deltat}{{\delta}}
In this appendix, we solve the Gross-Pitaevskii equation \eqref{eqMFproblem} and the Bogoliubov-de-Gennes problem \eqref{eqBgEV} perturbatively for weak tilt $\Delta$.
We use $UN/2$ as energy scale; in particular, we define the dimensionless Bogoliubov frequency $w=2\omega/UN$.
The dependence on the dimensionless parameter $y:=2J/UN$ is treated exactly.
We expand all quantities as $\mu = \mu^{(0)} + \deltat \mu^{(1)} + \deltat^2 \mu^{(2)} + \ldots$, where 
$\deltat = 2\Delta/(UN+2J)$ is the small parameter, which is the dimensionless smoothed tilt potential potential \cite{Sanchez-Palencia2006}.

With Eqn.~(\ref{eqn:Imbalance_exact}) and Eqn.~(\ref{muEq}), the perturbative solution of the meanfield problem \eqref{eqMFproblem} reads
\begin{align}\label{eqn:expansion-f}
 \begin{pmatrix}
  f_1\\ f_2
 \end{pmatrix}
 =
 \begin{pmatrix}
  f^{(0)} + \deltat f^{(1)} + \deltat^2 f^{(2)}+\ldots\\
  f^{(0)} - \deltat f^{(1)} + \deltat^2 f^{(2)}+\ldots
 \end{pmatrix}, \\
 f^{(0)} = 1/\sqrt{2} , \quad
 f^{(1)} = -f^{(0)}/2 , \quad
 f^{(2)} = -f^{(0)}/8. 
\end{align}

Next, we come to the perturbative solution of the Bogoliubov-de-Gennes equations \eqref{eqBgEV}.
The first orders of the matrices $d=2D/UN$ and $b=2B/UN$ read
\begin{align}
 d^{(0)} &= (1+y) \, \mathds{1} - y \sigma_x , & 
 b^{(0)} &= \mathds{1} , \\
 d^{(1)} &= -(1-y) \sigma_z , &
 b^{(1)} &=      - \sigma_z , \\
 d^{(2)} &= {y}/2 \, \mathds{1} , & 
 b^{(2)} &= 0  .
\end{align}
We observe that even orders commute with $\sigma_x$, whereas odd orders anti-commute.
This results in the following expansion of the Bogoliubov mode:
\begin{align}
 \omega &= \omega^{(0)} + \deltat^2 \omega^{(2)} + \ldots, \\
 \begin{pmatrix}
  u_1\\ u_2\\ v_1\\ v_2
 \end{pmatrix}
&=
 \begin{pmatrix}
 \phantom{-}u^{(0)} + \deltat\, u^{(1)} + \deltat^2 u^{(2)}+\ldots\\
          - u^{(0)} + \deltat\, u^{(1)} - \deltat^2 u^{(2)}+\ldots\\
 \phantom{-}v^{(0)} + \deltat\, v^{(1)} + \deltat^2 v^{(2)}+\ldots\\
          - v^{(0)} + \deltat\, v^{(1)} - \deltat^2 v^{(2)}+\ldots
 \end{pmatrix} . \label{eqn:expansion-uv}
\end{align}
At each order, the problem reduces to a $2\times 2$ problem, which is conveniently expressed in terms of the matrix
\begin{equation}
M_\pm = \begin{pmatrix}
  d_{11} \pm d_{12} & -(b_{11} \pm b_{12}) \\
  b_{11} \pm b_{12} & -(d_{11} \pm d_{12}) \\
\end{pmatrix}.
\end{equation}

The zeroth order consists in diagonalizing the matrix $M_-^{(0)}$, which yields
\begin{align}
 w^{(0)} &= 2 \sqrt{y(1+y)} , \label{eqDelta0-w} \\
 \begin{pmatrix}
  u^{(0)} \\
  v^{(0)}
 \end{pmatrix}
 &=\frac{1}{2\sqrt{(1 + 2y + w^{(0)})^2 - 1}}
 \begin{pmatrix}
  1 + 2y + w^{(0)} \\
  1
 \end{pmatrix} .\label{eqDelta0uv}
\end{align}

The first-order equation is of the form 
\begin{align}
\left[ M_+^{(0)}
-w^{(0)} \right]
 \begin{pmatrix}
  u^{(1)} \\
  v^{(1)}
 \end{pmatrix}
=
-M_-^{(1)}
 \begin{pmatrix}
  u^{(0)} \\
  v^{(0)}
 \end{pmatrix}
\end{align}
and is easily solved by inverting the matrix on the left hand side:
\begin{align}\label{eqn:1storder-uv}
 \begin{pmatrix}
  u^{(1)} \\
  v^{(1)}
 \end{pmatrix}
= \frac{1}{4y(1+y)}
\begin{smallpmatrix}
  y-(1-y)w^{(0)} & y + w^{(0)}\\
  y - w^{(0)}    & y+(1-y)w^{(0)}
 \end{smallpmatrix}
 \begin{pmatrix}
  u^{(0)} \\
  v^{(0)}
 \end{pmatrix}
\end{align}

The solution of the second order
\begin{align}
 \Bigl[M_-^{(0)} &- w^{(0)} \Bigr] 
 \begin{pmatrix}
  u^{(2)} \\
  v^{(2)}
 \end{pmatrix}\label{eq:PerturbativeDelta2}\\
 &= -\left\lbrace
  \left[M_-^{(2)} - w^{(2)} \right] 
 \begin{pmatrix}
  u^{(0)} \\
  v^{(0)}
 \end{pmatrix}
 + M_+^{(1)}
 \begin{pmatrix}
  u^{(1)} \\
  v^{(1)}
 \end{pmatrix}
 \right\rbrace  \nonumber
\end{align}
is less trivial, because the matrix on the left hand side is not invertible,
since its eigenvectors $(u^{(0)},v^{(0)})^t$ and $(v^{(0)},u^{(0)})^t$ have eigenvalues $0$ and $-2w^{(0)}$, respectively.
In order to solve \prettyref{eq:PerturbativeDelta2}, we expand the second order in terms of the zeroth order:
$(u^{(2)},v^{(2)}) = a^{(2)} (u^{(0)},v^{(0)}) +  c^{(2)} (v^{(0)},u^{(0)})$.
Then, we solve for the three unknowns $w^{(2)}$, $a^{(2)}$, and $c^{(2)}$ 
by multiplying Eqn.~(\ref{eq:PerturbativeDelta2}) from the left with $(u^{(0)},-v^{(0)})$, $(v^{(0)},-u^{(0)})$, and by employing the normalization condition to second order:
\begin{subequations}\label{eqn:2ndorder-uv}
\begin{align}
 w^{(2)} 
  &= 4y (2y-1) u^{(0)} v^{(0)} \label{eqn:2ndorder-w}\\
 c^{(2)} 
 &= \frac{-1}{8(1+y)} , \quad 
 a^{(2)} = - \left\{ |u^{(1)}|^2 - |v^{(1)}|^2 \right\} .
\end{align}
\end{subequations}
Remarkably, the renormalization of the Bogoliubov frequency $\omega = w UN/2$ can be either positive or negative, \prettyref{eqn:2ndorder-w}.

\paragraph*{Condensate depletion.}
Finally, we combine the previous results \eqref{eqn:expansion-uv}, \eqref{eqDelta0uv}, \eqref{eqn:1storder-uv}, and \eqref{eqn:2ndorder-uv} to 
compute the $\overline{v}_j$, which are needed for the depletion \eqref{eqn:CondepFetter}, up to second order.
$\overline{u}_j$ and $\overline{v}_j$ are expanded the same way as $u_j$ and $v_j$ in \prettyref{eqn:expansion-uv}, with $\overline{v}^{(0)} = v^{(0)}$, $\overline{v}^{(1)} = v^{(1)} - \xi^{(1)} f^{(0)}$, $\overline{v}^{(2)} = v^{(2)} - \xi^{(1)} f^{(1)}$, with $\xi^{(1)} =  2(f^{(0)}v^{(1)} + v^{(0)}f^{(1)})$.
We arrive at
\begin{align}
 \delta N^{(0)} &= 2 |\overline{v}^{(0)}|^2 = \frac{1}{2 w^{(0)} (1+2y + w^{(0)})} , \\
 \delta N^{(2)} 
 &= -\frac{3}{8}\frac{1}{(1+y)w^{(0)}}.
\end{align}
Thus, the initial change of the depletion 
is negative for all $y=2J/UN$ and scales quadratically with the tilt. 

\paragraph*{Zero mode.}
In order to transform the Hamiltonian \eqref{eqn:BgHamiltonian} from fluctuations $\delta \hat a_j$ and $\delta \hat a_j^\dagger$ to the Bogoliubov quasiparticle $\hat \beta$, $\hat \beta^\dagger$ and the self adjoined zero mode $\hat P = \sum_j f_j (\delta \hat a_j + \delta \hat a_j^\dagger)$, we also need the conjugate variable $\hat Q$, which is determined by 
\begin{align}\label{eqn:conditions-Q}
[\hat Q,\hat P]&=i, &
\hat Q^\dagger &= \hat Q, &
[\hat Q,\hat \beta]&=0 .
\end{align}
This is achieved by the ansatz
$\hat Q = \sum_j \gamma_j ( i \delta \hat a_j - i \delta \hat a_j^\dagger)$, where the amplitudes $\gamma_j$ are expanded in the same way in $\deltat$ as the amplitudes $f_j$ in \prettyref{eqn:expansion-f}.
From the conditions \eqref{eqn:conditions-Q}, we determine
\begin{align}
 \gamma^{(0)} &= \frac{1}{4f^{(0)}}, \qquad
 \gamma^{(1)}  = -\frac{u^{(1)} + v^{(1)}}{u^{(0)} + v^{(0)}} \, \gamma^{(0)} , \\
 \gamma^{(2)} &= -{4\gamma^{(0)}} \bigl[ \gamma^{(0)} f^{(2)} + \gamma^{(1)} f^{(1)} \bigr] .
\end{align}


Then, we can express the operators $\delta \hat a_j$ and $\delta \hat a_j^\dagger$ in terms of $\hat \beta$, $\hat \beta^\dagger$, $\hat P$, and $\hat Q$, which indeed brings the Hamiltonian \eqref{eqn:BgHamiltonian} to the form given in \prettyref{eqn:BgHamiltonian_diagonal}. We have already determined the Bogoliubov frequency $\omega = UN w/2$ above in Eqs.\ \eqref{eqDelta0-w} and \eqref{eqn:2ndorder-w}. Similarly, we determine the inverse mass parameter $\alpha$:
\begin{align}
 \alpha 
 &
  = UN\Bigl[ 1 + \frac{8\Delta^2 J}{(UN+2J)^3} + \ldots \Bigr] 
, \label{eqn:2ndorder-alpha} 
\end{align}
which is positive for all values of $J/UN$.
The numerical solution of the Bogoliubov-de-Gennes equation shows that $\alpha$ tends to $2UN$ for strong tilt $\Delta$.

%
%
%
%
%
%
%
%
%
%
\end{document}